\documentstyle[12pt]{article}
\include{figures}
\begin{document}
\newcommand{\kslash}{\mbox{$\displaystyle\not\mkern-4mu k$}}
\newcommand{\Dslash}{\mbox{$\displaystyle\not\mkern-4mu D$}}
\mbox{ }\hfill{\normalsize UCT-TP 196/92}\\
\mbox{ }\hfill{\normalsize BI-TP 93/36}\\
%
\begin{center}
{\Large \bf Dissipative Processes in an\\[.5cm]
 Expanding Massive Gluon Gas}\\
\vspace{1cm}
{\footnotesize{\bf J. Cleymans}\footnote{Department of Physics,
University of Cape Town, Rondebosch 7700, South Africa},
{\bf S.V. Ilyin}\footnote{Department of Physics,
Saratov State University, 410071 Saratov, Russia},
{\bf S.A. Smolyansky$^2$, G.M. Zinovjev}\footnote{Bogolubov
Institut for Theoretical Physics, Academy of
Science of Ukraine, 252143 Kiev, Ukraine.}}\\[.5cm]
\end{center}
\date{}
\vspace{.5cm}
\begin{abstract}
The temperature dependence of the kinetic coefficients
is obtained in the
non-perturbative region with the help of Green-Kubo-type
formulae in the model of massive gluon gas motivated by numerical
results from
simulations of lattice QCD. The
entropy production rate is estimated using scaling hydrodynamics.
It is shown that the
increase in the viscosity coefficients leads to entropy generation
in heavy ion collision processes which could be big,
especially for
temperatures  close to the critical one.
\end{abstract}
\newpage
\indent\indent
Discussions of the forthcoming projects for ultrarelativistic
heavy ion  collisions at RHIC and LHC require, with necessity,
comprehensive estimates of the space-time development in
those  reactions.
 Nowadays,  it is commonly believed that
 central heavy ion
 collisions pass through several stages. Describing these
in terms of equilibrium processes when all memory of prior
history has been destroyed, it is natural to
consider the entropy to set up the
equilibration time scale and to separate a preequilibrium stage. In
fact, that stage is producing not only the maximum entropy attained
[1,2], but a lot of uncertainties in the initial and boundary
conditions for the subsequent hydrodynamic expansion.
It has usually been
taken as isentropic, however it became clear that this stage may
be complicated by the dissipative processes generating  entropy,
together with a possible phase transition (or
transitions) when it happens to be of first order [3,4,5]. The
aim of the present study is therefore to explore the entropy
generation just at this stage although we understand that the
 freeze-out stage, where the system is made up of
free-streaming final particles,
could also add to the entropy [6,7].\\
\indent In order to estimate properly the dissipative effects
as well as the dynamics [8] of the QCD phase
transition we need to know
the behaviour of the kinetic coefficients (KC) over
a wide range of temperatures, including the
ones close to the phase transition point where
non-perturbative effects are dominant.
 The calculations done previously suffered from the unjustified
extrapolation
of the asymptotic behaviour,
found perturbatively, to the critical region [9,10,11]. The recent
estimate of the shear viscosity coefficient, using a  model for the
contribution of the nonperturbative region,
has demonstrated that the behaviour in the critical region
is very different from the  standard $T^{3}$ one and the amount of
entropy generated in the region close to this temperature appears
to be substantial [12,13]. \\
\indent In those calculations
we have exploited the so-called
momentum ``cut-off model''
motivated by a special analysis of the numerical results
of lattice Yang-Mills field thermodynamics [14,15]. In
spite of the fact of providing a good fit
to the data, this model  is very much
phenomenological
and doesn't interpret why the
low-momentum modes are removed nor does it
answer  questions in terms of
conventional conceptions of spin systems which are very indicative
and conclusive at least for the Monte Carlo analysis of lattice pure
gluodynamics.
Here we are dealing with the model of an ideal gas of excitations
where
the effect of interactions in the plasma is provided by the
temperature
dependence of their effective  mass [16].  In a sense it is rather
similar
to the ``cut-off model'' as now the medium (plasma) properties
suppress
low momentum excitations too,  since for $M(T)$
increasing
$\exp(-\sqrt{M^2+p^2}/T)\rightarrow \exp(-M/T)$ at small momenta
$p\rightarrow 0$. But it has more advantages as was argued in a
recent
analysis [17], the most important of them is that it leads to a
perfectly
detailed description of high precision SU(2)
pure gluodynamics lattice data just in the weak
coupling limit. \\
\indent However, as to the KC calculations, this
model is more involved as in what
follows we need to deal with massive scalar $\lambda\phi^4$ theory.
By the way,
this latter
approximation is also in line with the model of ``massive'' gluons
where these waves are nevertheless considered to be only transversal
[18].
Combining our calculations of KC's in $\lambda\phi^4$-theory with the
temperature dependence of the gluon mass extracted from lattice Monte
Carlo data, in particular for the mass gap in pure gluodynamics [19],
we
are able to estimate the entropy generated based on linear
hydrodynamics
and to explore the applicability of this approach to the
evolution of gluon systems.
\\
\indent As our basic hydrodynamical equation we take
\begin{equation}
\frac{d \varepsilon}{d \tau}+\frac{\varepsilon+p}{\tau}-
\frac{\chi}{\tau^2}=0.
\end{equation}
with Bjorken initial and boundary conditions [20].
We fix the equation of state ($p=p(\tau_{0}),
\varepsilon=\varepsilon(\tau_0)$ are the initial pressure and energy
density respectively), and taking the initial conditions at a time
$\tau_0\sim 1$ fm.
The dissipative term in Eq.(1) contains the factor $\chi
=(4/3\eta_s+\eta_v)$
with $\eta_s$and $\eta_v$ as transport coefficients of shear and bulk
viscosities. We should as well take into account that
$\tau >\chi/(\varepsilon+p)$, otherwise we  deal with the unrealistic
picture of gluon gas contraction (see [5,11]).\\
\indent The total entropy of the system is defined as [5,10]
\begin{equation}
S=\,\int d \sigma^{\mu}s^{\mu}=\int d\,y\,s(\tau)\,\tau\,\,,
\end{equation}
where $s^{\mu}=s\,u^{\mu},\,s(\tau)=[\varepsilon(\tau)+p(\tau)]/T$ \,
is the local entropy density and
$y$ is the
hydrodynamic rapidity (tanh\,$y = x/t$). Then Eqs.(1) and (2) give us
a
simple formula to estimate the entropy production in
expanding
 gluon gas
\begin{equation}
\frac{dS}{dy}=\int \frac{d\tau\,\chi (\tau)}{\tau\,T(\tau)}
\end{equation}
here $\chi(\tau)=\chi[T(\tau)],~~~ T(\tau)$ being the solution of
Eq.(1).
In order to solve Eq.(2) we need the temperature dependence of  the
KC's
and the equation of in the whole temperature interval, including the
nonperturbative critical region.

For pressure and energy density calculations we use
the familiar expressions [21]
\begin{equation}
p(T)=\frac{g}{6\pi^2}\,\int\limits_{0}^{\infty}\,dk\,\frac{k^4}{E(k)}\
\end{equation}
\begin{equation}
\varepsilon(T)=\frac{g}{2\pi^2}\,
\int\limits_{0}^{\infty}\,dk\,k^2\,E(k)\,n(E(k)),
\end{equation}
where $E(k)=\sqrt{k^2+m_{g}^{2}}\,$ is the relativistic energy;
$n(x)=[\exp (\beta x)-1]^{-1}$ is Bose's distribution function;
$\beta$ is the  inverse temperature and $g$ is the degeneracy factor.
In the ultrarelativistic case $(T\,>>m_{g})$ it leads to the
Stephan-Boltzmann (SB) law (see Fig.(1)).
However, if the temperature
is close to $T_c$,
the gluon mass $m_g$ is increasing, (at least for SU(2)-gluodynamics)
and becomes too
large to be negligible
and Bose's distribution function may be replaced by Boltzmann's
$(n(x)=\exp (-\beta x))$.
Then eqs (6),(7) may  be taken in the following form
\begin{equation}
p(T)=\frac{g\,T^4}{6\pi^2}\,\int\limits_{\alpha}^{\infty}\,
dz\,[z^2-\alpha^2]^{3/2}\,e^{-z}
=\frac{g\,T^4}{2\pi^2}\,\alpha^2\,K_2 (\alpha),
\end{equation}
\begin{equation}
\varepsilon(T)=\frac{g\,T^4}{2\pi^2}\,\int\limits_{\alpha}^{\infty}\,
dz\,z^2\,\sqrt{z^2-\alpha^2}\,e^{-z}=\frac{g\,T^4}{2\pi^2}\,\alpha^2\,
\left(3\,K_2 (\alpha)+\alpha \,K_1 (\alpha)\right),
\end{equation}
where $K_i(\alpha)$ are modified Bessel function and  $\alpha\equiv
m_g\beta$.
It is a well-known fact that  the asymptotic expansion of the modified
Bessel functions is given by
$K(\alpha)\sim \exp (-\alpha)\sqrt{\pi/2\alpha}$
for large $\alpha$, i.e. for
$T\sim T_c$, this way both pressure and energy
become small close to $T_c$. In the
case when $\alpha\rightarrow 0$ we recover the Stephan-Boltzmann
law.
This temperature dependence
can  fit even  the SU(3) Monte-Carlo lattice data with constant gluon
mass
[18] or including finite jump of mass around $T_c$. As to
SU(2)-gluodynamics, for which a much more elaborated  Monte Carlo
analysis
exists [17], we used the following parametrization
$$
m_g=m_0T_c\left( {T_c\over T-T_c} \right)^q
$$
with $m_0=1.83$ and $q=0.4$.
An important difference just reflects our
understanding in behaviours of  first- and second order phase
transitions
as seen in lattice Monte-Carlo simulations [19].
\indent To
calculate the KC's of shear and bulk viscosities
we use well-known relations obtained
within a formalism based on Kubo - type
formulae for the $\lambda\phi^4$ - thermofield theory [10,11] (The
analogous
expression for $\eta_s$ was also obtained for the case of vector
fields
[22])
\begin{equation}
\eta_s=\frac{\beta}{15}\,I_{2,1},
\end{equation}
\begin{equation}
\eta_v=\frac{\beta}{9}\biggl\{I_{2,1}-6c^{\,2}_{s} \,I_{1,0}
+9 \,c^{\,4}_{s}I_{0,-1}\biggr\},
\end{equation}
where $c^{\,2}_{s}=\partial p/\partial\varepsilon$
is the square of sound velocity, and the integrals $I_{m,n}$ are
defined as
\begin{equation}
I_{m,n}=2\,\int\frac{d^3{\tilde p}\,\,\,p^{2m}}{E^{2n}(p)\Gamma({\bf
p})}
\,\,n(p)\,[\,1+n(p)],
\end{equation}
here $\Gamma ({\bf p})$ is the
damping rate   of quasiparticle excitation
(it is assumed that $\Gamma \beta <<1$). For the scalar theory in one-
loop
approximation is
\begin{equation}
\Gamma({\bf p})=\frac {\lambda^2(2\pi )^4}{24E({\bf p})n({\bf p})}
\int d^3{\tilde p}_1d^3{\tilde p}_2
d^3{\tilde p}_3\delta (p+p_1-p_2-p_3)(1+n_1)\,n_2\,n_3
\end{equation}
with the following designations \,$d^3{\tilde p}=
[2(2\pi )^3E({\bf p})]^{-1}d^3p$, in both Eqs.(11) and (12),\,
$n_i=[\exp ({\beta}E({\bf p}_i))-1]^{-1}$.\\
\noindent The gluon gas
viscosities may be obtained from Eqs.(8)-(11) by the standard
procedure of changing [11,23]
\begin{equation}
\lambda^2\longrightarrow\,c^*32\pi^2\alpha^2_s\ln\alpha^{-1}_s,
\quad c^*=20\div 60,
\end{equation}
where (for the value of $c^*$, see also [24])
\begin{equation}
\alpha_s=6\pi\Bigl[11/2\,N\ln (M^2/\Lambda^2)\Bigl]^{-1},
\end{equation}
and $N$ is a number of colours, $M^2=\frac{4}{3}<p^2>$ and $<p^2>$
is the thermodynamicaly averaged squared momentum of the gluon field
[21].
The degeneracy factor $g=(d-1)(N^2-1)$ is absent in final result,
for the numerator of Eq.(12) must contain it as well as the damping
rate  in the denominator. This phenomenological estimate can also be
justified by the fact that in the lowest order of interaction the
cross
sections for gluons and scalar particles have
a similar momentum dependence.
This procedure (Eq.(13)) is fair only for small
enough interaction constant.
The model under consideration brings us to the following temperature
dependence for the $M^2$ factor from Eq.(15)
\begin{equation}
M^2=\frac{3}{4}\,\frac{\int d^{3}p\,p^2\,n(E(p)}{\int
d^{3}p\,n(E(p))}=
4\,T^2\,\alpha\,\frac{K_3 (\alpha)}{K_2 (\alpha)}.
\end{equation}
In the ultrarelativistic case where $\alpha\rightarrow 0$ we obtain
the
conventional result $M\sim 4T$. When $T\rightarrow T_c$, and
$\alpha\rightarrow\infty$ we have $M^2\sim 4T^2 \alpha$.
This means that the
coupling constant $\alpha_s$ remains still small even when
the temperature is close to $T_c$. This lucky fact has been met
already
 in the so-called cut-off model [14,15] that interprets Monte-Carlo
data as
well as our model does. It can be explained evidently by the fact that
the application of both gluodynamic models takes the contribution
of long wavelength excitations away.

We now proceed to calculate the damping rate  (Eq.(11)).
After some straightforward
 integrations using the $\delta$ - function we have
\begin{equation}
\Gamma({\bf p})=\frac {\lambda^2(2\pi )^{-4}}{192E(p)n(E(p))}
\int d^3\,p_{1}\frac{(1+n_1)}{E({\bf p}_1)}\,I_1,
\end{equation}
where
$$I_1=2\pi\beta^{-1}\int\limits_{\alpha}^{\beta\Omega}\,dy\,\exp (-
\beta\Omega)
\,\Theta(z_{0} (y)-1)\,\Theta(z_{0} (y)+1),$$
\begin{equation}
z_{0} (y)=\frac{K^2-\Omega^2+2\Omega y}{2K\sqrt{y^2-m_g^2}},
\end{equation}
here we need to solve an inequality
\begin{equation}
-1\leq z_{0} (y)\leq 1,
\end{equation}
where
\begin{equation}
\Omega=\sqrt{p_1^2+m_g^2}+\sqrt{p^2+m_g^2}
\end{equation}
 and
$K=\sqrt{p^2+p_1^2+2p\,p_1\,\cos (\theta)}$, $\theta$ being the angle
between $\bf p$ and $\bf p_1$.
It then leads us to the expression
\begin{equation}
y_{-}\leq y\leq y_{+};~~y_{\pm}=\frac{\Omega}{2} \pm
\frac{K\Lambda}{2};~~
\Lambda=\sqrt{1-\frac{4m_g^2}{\Omega^2-K^2}},
\end{equation}
with the values of $\Omega $ and $K$ satisfying the inequality
\begin{equation}
\Omega^2-K^2\,>\,4m_g^2.
\end{equation}
After analysis of  Eqs. (19) and (20) we obtain for Eq.(15)
\begin{equation}
\Gamma({\bf p})=\frac{\lambda^2 (2\pi)^{-3}}{\beta^{2}192E(p)}
\int\limits_{0}^{\infty} dx\frac{x^{2}[1+\exp(-\sqrt {x^2+\alpha^2})]}
{\exp(-\sqrt {x^2+\alpha^2})\sqrt {x^2+\alpha^2}}
\int\limits_{0}^{\pi}d\theta\sin(\theta) K\Lambda.
\end{equation}

These relations are  complicated to integrate, in order to get an
approximate
analytical form we can
calculate the angle integral from Eq. (21) in the $p=0$ case.
This gives  a minimal value for the damping rate  and a
maximum one for the kinetic
coefficients. Here we also take into account that the main
contribution in
Eqs.(8)-(10) is connected with long wavelenght excitations. Thus, in
this
approximation the damping rate  looks as
\begin{equation}
\Gamma(p)=\frac{\lambda^2(2\pi)^{-3}\,\beta^{-2}}{96\,E(p)}
\int\limits_{\alpha}^{\infty}\,dz\,(z-\alpha) \,
\sqrt{z^2-\alpha^2}\left(e^{-z}+e^{-2z}\right).
\end{equation}

It is evident that the momentum  dependence is very simple in this
equation.
This allows us to represent the
integral $I_{m,n}$ from Eq.(10) in the
following form
\begin{equation}
I_{m,n}=\frac{T^{2(n-m)-2}}{2\,\pi^2\,\tilde\Gamma(T)}\,J_{m,n}
\end{equation}
with
\begin{equation}
J_{m,n}=\int\limits^{\infty}_{\alpha}dz\,z^{2-2n}\,(z^2-\alpha^2)^
{\frac{2m+1}{2}}[\,\exp (-z)+\exp (-2z)]
\end{equation}
After  integration we obtain, for some specific cases
\begin{equation}
J_{2,1}=\frac{15}{8}\,[(2\alpha)^{3}\,K_3
(\alpha)+\alpha^{3}\,K_{3}(2\alpha)],
\end{equation}
\begin{equation}
J_{1,0}=3\alpha^3\,[\frac{5}{8}\,K_3 (2\alpha)+5\,K_3 (\alpha)+
\frac{\alpha}{4}\,K_2 (2\alpha)+\alpha\,K_2 (\alpha)],
\end{equation}
$$J_{0,-1}=\alpha^2 \,[60\,K_2 (\alpha)+\frac{15}{4}\,K_2 (2\alpha)+
27\,\alpha\,K_1 (\alpha)+\frac{27}{8}\,K_1 (2\alpha)+$$
\begin{equation}
+6\alpha^2\,K_0 (\alpha)+\frac{3}{2}\alpha^2 K_0 (2\alpha)+
\alpha^3 K_1 (\alpha)+
\frac{\alpha^3}{2}\,K_1 (2\alpha).
\end{equation}

The result for the damping increment can be written as
\begin{eqnarray}
\tilde\Gamma(T)&=&E(p)\,T^2\,\Gamma(p)\nonumber\\
               &=&\frac{\lambda^2\,\pi^{-3}}{3\,2^9}\,
\alpha^2(T)\left[2K_2(\alpha)-2K_1(\alpha)+K_2(2\alpha)-
K_1(2\alpha)\right]
\end{eqnarray}
\indent For the calculation of
the bulk viscosity we also need to know the expression for the
velocity of sound. With the help of Eqs. (6) and (7) we obtain
$$c^{\,2}_{s}=\frac{1}{3+\Delta}$$
with
\begin{equation}
\Delta=\,\alpha\,\frac{4K_1(\alpha)
+T\,\frac{d\alpha}{dT}\,[\,K_2 (\alpha)-2K_0 (\alpha)]}
{4K_2(\alpha)-T\frac{d\alpha}{dT}\,K_1 (\alpha)}
\end{equation}
Here we see that, for $\alpha$ small,  $\Delta\sim\alpha$, and
the sound velocity $c_s$ vanishes. In the $\alpha\rightarrow 0$ case
Eq.(28) gives the common value of $c_s^2=1/3,\,\,\,\Delta\rightarrow
0$
(see Fig.(2)).

Eqs.(24-28) together with  Eqs.(8,9,10,12-14)
determine the temperature
dependence of the KC of the gluon gas.
The  asymptotic behaviour
$\alpha\rightarrow 0$ gives  results analogous to Ref. [11] that is
somewhat  large, for, here we have
a maximal estimate of KC (see below Eq.(21)).
The temperature dependences of the KC are
 presented in Fig.(3). It is evident
that the KC increase considerably  in the
temperature region close to $T_c$. This leads to a big deviation
of the solution of Eq. (1) from the scaling one, and to the
so-called critical delay of the
evolution of the system  near $T_c$.

The rate   of entropy production
as a function of proper time is depicted in Fig. (5).
It is evident that the entropy of the system increases rapidly.
The analogous calculations for entropy production rate for $T^3$-
dependency
of KC brings to approximately 20 per cent increasing of entropy in
gluon gas
cooling process.

The authors are aware of the fact that the obtained results are
model dependent. The increase of the kinetic coefficients
 makes the application of linear
hydrodynamics for the description of quark gluon gas questionable.
Evidently, taking into account other dissipative mechanisms
will lead to finite values for the KC and diminish somehow the
entropy production. However the results obtained
indicate that a realistic picture of the evolution of the
system under consideration can differ a lot from the scaling one,
especially in the phase transition region and demands thorough
examination.
\newpage
\begin{center}
Figure Captions\\
\end{center}
\vspace{0.1cm}
\begin{description}
\item[Figure 1.] Dependence of the scaled energy density and pressure
on the
temperature.
\item[Figure 2.] Dependence of the speed of sound on the
temperature.
\item[Figure 3.] Evolution of the temperature with proper time.
\item[Figure 4.] Dependence of the shear and bulk viscosities on the
temperature.
\item[Figure 5.] Entropy generation as a function of proper time.
\end{description}
\newpage
\begin{center}
\bf References \\
\end{center}
\vspace{0.1cm}
\begin{description}
\item[1.] K. Geiger, Phys. Rev. D46, (1992) 4986.
\item[2.] D. Seibert,  Phys. Rev. Lett., 67 (1991) 12.
\item[3.] M. Gyulassy, T. Matsui, Phys. Rev. D29 (1984) 419.
\item[4.] Y. Akase, S. Date, M. Mizutani, Sh. Muroya, M. Namiki, M.
Yasuda,
 Prog. Theor. Phys. 82 (1989) 591.
\item[5.] Y. Akase, M. Mizutani, Sh. Muroya, M. Yasuda,
Prog. Theor. Phys. 85 (1991) 305.
\item[6.] J. Cleymans, K. Redlich, H. Satz, E. Suhonen, Preprint CERN-
TH
6759/92.
\item[7.] P. L\'evai, B. Luk\'acs, J. Zim\'anyi,, J. Phys. G16 (1990)
1019;
 U. Heinz, J. Letessier, J. Rafelski, A. Tounsi, Preprint PAR-LPTHE
\item[8.] L. Csernai, J. Kapusta, Phys. Rev. Lett. 67  (1992) 737.
\item[9.] P. Danielewicz, M. Gyulassy, Phys. Rev. D31 (1985) 53.
\item[10.] A. Hosoya , K. Kajantie, Nucl. Phys. B250 (1985) 666.
\item[11.] R. Horsley, W. Schoenmaker ,  Nucl. Phys. B280 (1987) 716;
735.
\item[12.] S. Ilyin, O. Mogilevsky, S. Smolyansky, G. Zinovjev, Phys.
Lett.
296B (1992) 365.
\item[13.] J. Cleymans, S. Ilyin, S. Smolyansky, G. Zinovjev, Preprint
BI-TP 92-42.
\item[14.] F. Karsch, Zeit. f. Phys. C38 (1988) 147;
 M. Gorenstein, O. Mogilevsky, Zeit. f. Phys, C38 (1988) 161.
\item[15.] J. Engels, J. Fingberg, K. Redlich, H. Satz, H. Weber,
 Zeit. f. Phys. C42 (1989) 341;
M. Gorenstien, O. Mogilevsky, S. Mrowczynski, Phys. Lett. B246 (1990)
200.
\item[16.] T. Biro, Int. Journ. Mod.  Phys., E1 (1992) 39.
\item[17.] V. Goloviznin, H. Satz, Zeit. f. Phys. C57 (1993) 671.
\item[18.] T. Biro, P. L\'evai, B. M\"uller, Phys. Rev. D42 (1990)
3078.
\item[19.] A. Ukawa,  Nucl. Phys. (Proc. Suppl.) 17 (1990) 118.
\item[20.] J.D. Bjorken,  Phys. Rev. D27 (1983) 140.
\item[21.] J. Kapusta,  Nucl. Phys. B148 (1979) 461.
\item[22.] S. Ilyin, A. Panferov, Yu. Sinyukov,
Phys. Lett. B227 (1989) 455.
\item[23.] A. Hosoya, M. Sakagami, M. Takao,  Ann. Phys. 154 (1984)
229.
\item[24.] G. Baym, H. Monien, C. Pethick et al.,  Phys. Rev. Lett.
64,
(1990) 1867.
\end{description}
\end{document}